\documentclass[aps,prd,
12pt,
nofootinbib,showpacs,superscriptaddress,preprintnumbers,amsmath,amssymb,floatfix]{revtex4-2}

\usepackage[bookmarks, linktocpage, colorlinks = true, linkcolor = blue, urlcolor  = blue, citecolor = blue, anchorcolor = green, hyperindex = true, hyperfigures]{hyperref}

\usepackage{graphicx} 
\usepackage{amssymb}
\usepackage{dcolumn}
\usepackage{multirow}
\usepackage{amsmath,amssymb}
\usepackage{slashed}
\usepackage[usenames]{color}
\usepackage{float}
\usepackage{tabularx}

\usepackage{hhline,multirow,tabularx}

\newcommand{\Nf}{N_{f}}
\newcommand{\bp}{\boldsymbol{p}}

\newcommand{\bq}{\boldsymbol{q}}
\newcommand{\bqhat}{\hat{\boldsymbol{q}}}
\newcommand{\kF}{k_{\mathrm{F}}}
\newcommand{\mD}{m_{\mathrm{D}}}
\newcommand{\DE}{D^{\mathrm{E}}}
\newcommand{\DM}{D^{\mathrm{M}}}

\begin{document}

\preprint{N3AS-25-015, RIKEN-iTHEMS-Report-25}

\title{Classification of color superconductivity by one-gluon exchange helicity amplitudes and renormalization group equations}

\author{Yuki~Fujimoto}
\affiliation{Department of Physics, Niigata University, Ikarashi, Niigata 950-2181, Japan}
\affiliation{Department of Physics, University of California, Berkeley, CA 94720, USA}
\affiliation{RIKEN Center for Interdisciplinary Theoretical and Mathematical Sciences (iTHEMS), RIKEN, Wako 351-0198, Japan}

\date{\today}

\begin{abstract}
    Quark matter at high baryon density exhibits diverse pairing patterns classified by color, flavor, and angular momentum quantum numbers.
    We compute one-gluon exchange (OGE) helicity amplitudes and introduce a nonrelativistic classification of the pairing channel, justified by the channel decomposition in a Lorentz-noninvariant medium and the decoupling of renormalization group flows at leading order.
    We find the new attractive channel in OGE;
    the medium effects can render the vacuum-repulsive color $\boldsymbol{6}$ channel attractive in the spin-triplet sector.
    For color $\boldsymbol{\bar{3}}$ with antisymmetric flavor, the dominant pairing is ${}^1S_0$, while for symmetric flavor the same-helicity $^1P_1$ prevails over $^3S_1$, revising the conventional single-flavor picture.
    With a mismatch of the Fermi momenta, $^1S_0$ channel, leading to color-flavor locked or two-flavor color superconductor, remains most stable when the separation is small, and the color-spin locked pairing becomes favored as the mismatch gets large.
    We suggest there are possible quark-hadron continuity in certain cases as expected in the literature.
\end{abstract}

\maketitle

\section{Introduction}

The QCD phase diagram at finite quark chemical potential $\mu$, at which matter is expected to be a color superconductor, remains poorly understood (see, e.g., Refs.~\cite{Alford:2007xm,Fukushima:2010bq,Fukushima:2025ujk} for reviews).
This region is crucial for uncovering the properties of matter inside neutron star cores, which is a necessary input for gravitational wave observation from binary neutron star mergers, and is also actively being explored in terrestrial heavy-ion collision experiments.
The main obstacle to progress is the sign problem, which prevents nonperturbative Monte Carlo calculations of QCD on the lattice~\cite{Nagata:2021ugx}.

The main purpose of this paper is to demonstrate the classification of the most attractive channel in quark-quark scattering and the corresponding color-\hspace{0pt}superconducting pairing pattern arising from this scattering.
The object we focus on is the one-gluon exchange (OGE) helicity amplitude, which was previously considered in Ref.~\cite{Bailin:1983bm} (see, Refs.~\cite{Jacob:1959at,Chung:1971ri} for a review on the formalism);
however, this work was before the advent of hard thermal loops~\cite{Pisarski:1988vd,Braaten:1989mz}.
To take this development into account, we perform the calculation in the hard dense loop (HDL) effective theory.
We work in the weak-coupling regime so that everything is under control perturbatively.

In analogy with nonrelativistic systems, color-\hspace{0pt}superconducting states can be classified by the quantum numbers $(S,L,J)$, where $S$ is the total spin, $L$ the orbital angular momentum, and $J = L + S$ the total angular momentum.
Each channel can be labeled concisely by the spectroscopic term symbol ${}^{2S+1} L_J$.
This classification is justified by the fact that spin remains a well-defined quantum number in the amplitude relevant for Cooper pairing in finite-$\mu$ QCD.
Remarkably, this labeling reveals that attractive interactions can emerge even in color sextet ($\boldsymbol{6}_c$) channels, which is repulsive in the vacuum.
This is owing to the dominance in the chromomagnetic gluons over the chromoelectric gluons in the dense medium;
in the HDL effective theory, magnetic gluons are less screened compared to electric gluons~\cite{Son:1998uk}.
There are other exotic scenarios, such as Kohn-Luttinger mechanism, to power an attraction in this channel~\cite{Schafer:2006ue,Kumamoto:2024muw, Fujimoto:2025zfa}, but the attraction we find is much stronger than these attraction.

The key observation that lead to the classification above is as follows.
Conventionally, the classification of pairing channels in relativistic systems has been based solely on the total angular momentum 
$J$, since $S$ and $L$ are not separately conserved due to Lorentz transformation.
However, at finite density, Lorentz boost invariance is explicitly broken by the presence of the special rest frame of the dense medium, and rotational symmetry alone remains good symmetry.
As a result, spin configuration is frozen in the medium rest frame;
the total spin remains invariant under the residual rotational symmetry and thus regains its nonrelativistic character.

We note in passing that there is another example in dense QCD in which the lack of Lorentz invariance plays a pivotal role:
The counting rule of the Nambu-Goldstone modes becomes different from that in the system with Lorentz invariance~\cite{Miransky:2001tw, Schafer:2001bq, Watanabe:2012hr, Hidaka:2012ym}.

Within this classification by a quartet of the labels (color representation, flavor representation, helicity, ${}^{2S+1}L_J$), the pairing gaps in different channels evolve independently.
This decoupling arises because the renormalization group (RG) equations, which governs the BCS instability, are diagonal in these channels at leading order in perturbation theory~\cite{Evans:1998ek, Evans:1998nf,Schafer:1998na, Son:1998uk, Hsu:1999mp, Fujimoto:2025inprep2}.
One can therefore compute the pairing gap for each channel separately, and determine the most attractive channel for a given color-flavor configuration.

The earlier consideration with the same intent of classifying the color-\hspace{0pt}superconducting by color, flavor, and spin channels has been made in Refs.~\cite{Alford:2002rz,Alford:2005yy}, where they used the Nambu--Jona-Lasinio model with an asymmetry between the electric and magnetic interaction.
In this work, we perform the similar analysis, but from the QCD perspectives.

The physics motivation of such a classification is to set a boundary condition in the QCD phase diagram in the large-$\mu$ region.
At sufficiently high densities and low temperature, at which QCD becomes weakly coupled, the phase is expected to be the color-flavor locked (CFL) color superconductor~\cite{Alford:1998mk}.
In this phase, the strange quark participates symmetrically with the up and down quarks in Cooper pairing.
As one lowers the density, the vacuum may undergo the transition to another non-CFL phase with less symmetry, such as two-flavor color-\hspace{0pt}superconducting (2SC) phase.
Below the CFL density, there can appear variety of phases with various pairing patterns, but the phase diagram is not understood well (see Sec.~III in Ref.~\cite{Alford:2007xm}).

In fact, even in the weak-coupling regime, the phase diagram has not been settled to date~\cite{Fujimoto:2024pcd}.
In charge-neutral and beta-equilibrium system, there is a mismatch of the Fermi surface induced by a finite strange quark mass $m_s$.
This induces stress between the particles participating in the BCS pairing, and pulls apart the pairing states.
If the Fermi momentum splitting induced by $m_s$ is large compared to the pairing gap $\Delta_{\rm CFL}$, the CFL phase is no longer favored~\cite{1962ApPhL...1....7C,Clogston:1962zz,Rajagopal:2000ff, Alford:2001zr}.
By using perturbative running quark mass for $m_s$~\cite{Marquard:2015qpa} and the weak-coupling gap for $\Delta_{\mathrm{CFL}}$~\cite{Brown:1999aq,Brown:1999yd,Wang:2001aq}, one can find the CFL phase becoming unstable even in the weak-coupling regime for some value of $\mu \gtrsim 1\,\text{GeV}$, although there is still large uncertainty, particularly the one associated with the renormalization scale variation.

The phase structure of high-density quark matter may have a consequence for astrophysics.
Although color superconductivity does not drastically affect the bulk thermodynamic properties, such as equation of state, it can affect phenomena sensitive to the degrees of freedom on the Fermi surface.
The notable example is the neutrino emission from quark matter, which is related to the neutron star cooling.
If a color-\hspace{0pt}superconducting gap is of keV order, it can suppress the cooling of neutron stars, while the unpaired quarks can power the rapid cooling via the quark direct Urca process (see, however, Ref.~\cite{Alford:2024xfb} for the recent literature on the reconsideration of the Urca process).

In this work, we limit ourselves to the homogeneous condensate.
We do not consider crystalline phases~\cite{Alford:2000ze,Bowers:2002xr,Casalbuoni:2003wh,Anglani:2013gfu}, the unstable gapless phases~\cite{Shovkovy:2003uu,Giannakis:2004pf,Huang:2004am,Casalbuoni:2004tb,Fukushima:2005cm}, or thermal phase transition with an increasing temperature~\cite{Iida:2003cc,Iida:2004cj} here.
These aspects, while interesting, are beyond the scope of the present analysis.
Finally, although our results are based on one-loop calculations, we do not expect higher-order corrections to alter the qualitative picture;
the overall gap magnitudes may shift by at most an order of magnitude.

The paper is organized as follows.
In Sec.~\ref{sec:helicity}, we compute the helicity amplitude of the OGE interaction in a fixed total angular momentum.
In Sec.~\ref{sec:decompose}, we map each total angular momentum states to the canonical states labeled by the total spin and angular momentum.
In Sec.~\ref{sec:classify}, we classify the possible pairing channel for a given color and flavor channel.
In Sec.~\ref{sec:pattern}, we discuss the possible pairing patterns in the neutron-star condition, in which the beta equilibrium and charge neutrality conditions are imposed.
Finally, we conclude the paper in Sec.~\ref{sec:conclusions}.

\section{Quark-quark helicity amplitude in dense QCD}
\label{sec:helicity}

In the nonrelativistic case, the interaction potential between the fermions enters the gap equation.
In the relativistic case of color superconductivity, the corresponding object are the helicity amplitudes for quark-quark scattering near the Fermi surface.
In terms of quark representations of color $SU(3)$, the diquark can be enumerated as follows. Because each quark is a color triplet $\boldsymbol{3}$, the pair can form $\boldsymbol{3} \otimes \boldsymbol{3} = \boldsymbol{\bar{3}} \oplus \boldsymbol{6}$;
the color $\boldsymbol{\bar{3}}_c$ is antisymmetric, and $\boldsymbol{6}_c$ is symmetric.
It is widely believed that the $\boldsymbol{\bar{3}}$ channel is attractive and $\boldsymbol{6}$ channel is repulsive.
Contrary to the standard view, we demonstrate that attraction can even be present in the color $\boldsymbol{6}$ channel.

We consider a scattering of massless quarks on the Fermi surface in the center-of-momentum frame: $q_{\lambda_1}(\bp_1) + q_{\lambda_2}(-\bp_1) \to q_{\lambda_3}(\bp_3) + q_{\lambda_4}(-\bp_3)$, where $\lambda_i$ is helicity of the $i$-th quark, the magnitude of the momenta are set to the Fermi momentum $|\bp_1| = |\bp_3| = \mu$, and the scattering angle $\theta$ is defined as $\hat{\bp}_1 \cdot \hat{\bp}_3 = \cos\theta$.
The helicity conservation requires $\lambda_1 = \lambda_3$ and $\lambda_2 = \lambda_4$.

A standard BCS condensate is position-independent so that in the momentum space the pairing is between quarks with equal and opposite momentum.
Because Cooper pairs have zero total momentum in the rest frame of the medium, the only scattering amplitude that enters the pairing kernel is the one computed in that very frame.
Choosing the center-of-momentum frame is therefore not just a convenience; it is in practice the indispensable choice whenever boost symmetry is broken by the presence of the finite-density medium.

\begin{figure}
    \centering
    \includegraphics[width=.45\columnwidth]{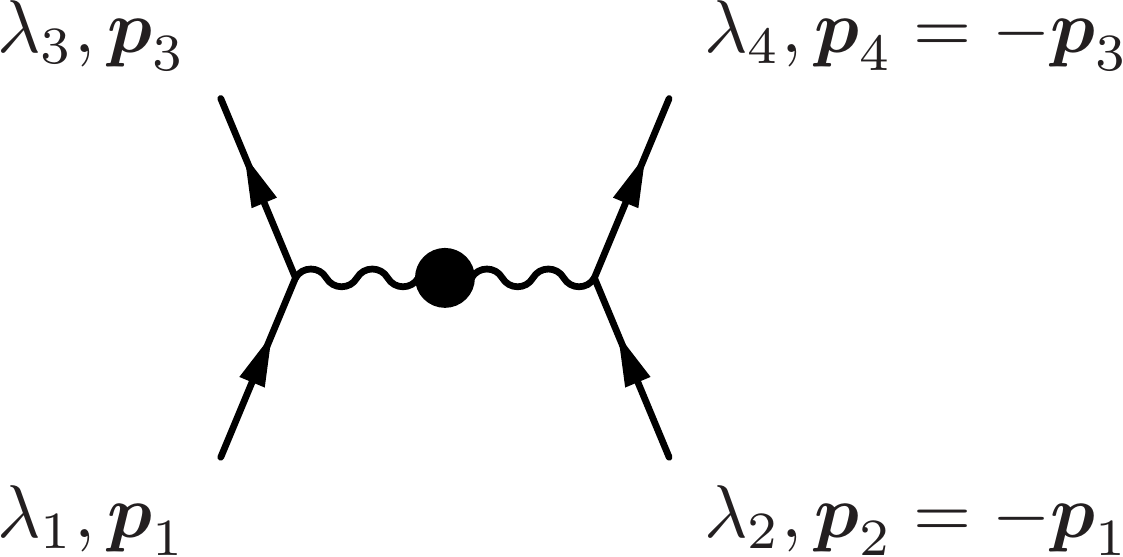}
    \caption{The tree-level helicity amplitude for quark-quark scattering via single gluon exchange.
    The heavy dot indicates the HDL-resummed propagator.}
    \label{fig:tree}
\end{figure}
We consider the amplitude corresponding to a Feynman diagram in Fig.~\ref{fig:tree}
\begin{align}
    i\mathcal{M}_{\lambda_1 \lambda_2; \lambda_3 \lambda_4}
    &= \bar{u}_{\lambda_3}(p_3) (ig \gamma^\mu t^a) u_{\lambda_1}(p_1)
    iD_{\mu\nu}(q)
    \bar{u}_{\lambda_4}(-p_3) (ig \gamma^\nu t^a)  u_{\lambda_2}(-p_1)\,,
\end{align}
where $g$ is the QCD coupling constant and $q = p_1 - p_3$ is the momentum transfer.
The gluon propagator $D_{\mu\nu}$ in the Landau gauge is
\begin{align}
    D_{\mu\nu} = \DE P_{\mu\nu}^{\mathrm{L}} + \DM P_{\mu\nu}^{\mathrm{T}}
\end{align}
where the transverse and longitudinal projectors $P_{\mu\nu}^{\mathrm{T}}$ and $P_{\mu\nu}^{\mathrm{L}}$ are given by
\begin{align}
    P_{\mu\nu}^{\mathrm{T}} &= -\delta_{\mu i} \delta_{\nu j} \left(\delta_{ij} - \bqhat_i \bqhat_j\right)\,,\\
    P_{\mu\nu}^{\mathrm{L}} &= g_{\mu\nu} - \frac{q_\mu q_\nu}{q^2} - P_{\mu\nu}^{\mathrm{T}} \simeq \delta_{\mu0} \delta_{\nu 0}\,.
\end{align}
The superscripts E and M correspond to the chromoelectric and magnetic gluon propagators, respectively.
Since the Lorentz invariance is lost at finite density, $\DE$ and $\DM$ does not have to be identical to each other.
We employ the resummed propagator in the HDL effective theory
\begin{align}
    D^{\mathrm{E/M}} &= - \frac{1}{q_0^2 - \boldsymbol{q}^2 - \Pi^{\mathrm{E/M}}}\,,
    \label{eq:gluonprop}
\end{align}
where $\Pi^{\mathrm{E/M}}$ is the electric (longitudinal) / magnetic (transverse) part of the gluon self-energy.
For $q_0 \ll |\bq| \to 0$ and to leading order in perturbation theory~\cite{Bellac:2011kqa, Laine:2016hma}
\begin{align}
    \Pi^{\mathrm{E}} &\simeq \mD^2\,,\\
    \Pi^{\mathrm{M}} &\simeq -i \frac{\pi}4 \mD^2\frac{q_0}{|\bq|}\,,
\end{align}
where $\mD^2 = \Nf g^2 \mu^2 / (2\pi^2)$ is Debye screening mass.
The magnetic interaction is not screened statically, but there is a dynamical screening with a frequency-dependent cutoff $q_c = (\pi \mD^2 q_0 /4)^{1/3}$.

The amplitude for quarks with the same helicities is
\begin{align}
    \mathcal{M}_{++;++}
    &= c_R g^2 \left[ \DE \cos^2 \frac{\theta}2 + \DM \left(\cos^2 \frac{\theta}2 + 2 \sin^2 \frac{\theta}2\right) \right]\,,\notag\\
    &= \frac{c_R g^2}{2} \left[ \DE + 3 \DM + \left(\DE - \DM\right) \cos\theta\right]\,,
\end{align}
where the subscript $++$ refers to the incoming quarks having the same helicity $\lambda_1 = \lambda_2 = +1/2$, and $c_R$ is the color factor for the representation $R$:
\begin{align}
    c_R = \begin{cases}
        -\frac23 & (R=\boldsymbol{\bar{3}})\\
        \frac13 & (R=\boldsymbol{6})\,.
    \end{cases}
    \label{eq:cR}
\end{align}
For the opposite helicities,
\begin{align}
    \mathcal{M}_{+-;+-}
    &= c_R g^2 \left( \DE \cos^2 \frac{\theta}2 + \DM \cos^2 \frac{\theta}2 \right)\,,\notag\\
    &= \frac{c_R g^2}{2} \left(\DE + \DM\right) \left(1 + \cos\theta\right) \,,
\end{align}
where the subscript $+-$ refers to the opposite helicity $\lambda_1 = +1/2, \lambda_2=-1/2$.

Now, we expand the amplitude in partial waves of the total angular momentum $J$, following the method by Jacob and Wick~\cite{Jacob:1959at,Chung:1971ri}.
In the relativistic case, the Wigner rotation has to be taken into account, so the expansion is in terms of the Wigner $d$-matrix $d^J_{\lambda \lambda'}(\theta)$ instead of the Legendre polynomial $P_J(\cos\theta)$.
\begin{align}
    \mathcal{M}_{\lambda_1\lambda_2;\lambda_3\lambda_4}(\theta) = \sum_J (2J+1) \mathcal{H}^J_{\lambda_1 \lambda_2} d^J_{\lambda \lambda'}(\theta) \,,
\end{align}
where $\lambda= \lambda_1 - \lambda_2$ and $\lambda' = \lambda_3 - \lambda_4$.
Inversely, from the orthogonality of the Wigner $d$-matrix, the helicity amplitude can be expressed as
\begin{align}
    \mathcal{H}^J_{\lambda_1 \lambda_2} = \frac12 \int_{0}^{\pi} d\theta\, \sin\theta\, d^J_{\lambda \lambda'}(\theta) \mathcal{M}_{\lambda_1 \lambda_2; \lambda_3 \lambda_4} (\theta)\,.
\end{align}

The OGE helicity amplitude in the $++$ channel is
\begin{align}
    \mathcal{H}^{J}_{++}
    &=  \frac{c_R g^2}{4} \int_{0}^{\pi} d\theta\, \sin\theta\, P_J(\cos\theta)
    \left[ \DE + 3 \DM + \left(\DE - \DM\right) \cos\theta\right]\,.
\end{align}
where we used the relation
\begin{align}
    d^J_{00}(\theta) &= P_J(\cos\theta)\,.
\end{align}
By performing the integral, we obtain
\begin{align}
    \mathcal{H}^{J}_{++}
    &= \frac{c_R g^2}{2} \left[D_J^{\mathrm{E}} + 3D_J^{\mathrm{M}} + \frac{J}{2J+1}\left(D_{J-1}^{\mathrm{E}} - D_{J-1}^{\mathrm{M}}\right)
    + \frac{J+1}{2J+1}\left(D_{J+1}^{\mathrm{E}} - D_{J+1}^{\mathrm{M}}\right)\right]\,,
    \label{eq:HJ++}
\end{align}
where $D_L^{\mathrm{E/M}}$ is the partial wave expansion of $D^{\mathrm{E/M}}(\theta)$:
\begin{align}
    D^{\mathrm{E/M}}(\theta) = \sum_L (2L+1) D_L^{\mathrm{E/M}} P_L(\cos\theta)\,,
    \label{eq:DL}
\end{align}
or, inversely, $D_L^{\mathrm{E/M}} = \frac12 \int_0^{\pi} d\theta\, \sin\theta\, P_L(\cos\theta) D^{\mathrm{E/M}}(\theta)
$.
In the derivation, we used the relation
\begin{align}
    x P_J(x) &= \frac{J}{2J+1} P_{J-1}(x) + \frac{J+1}{2J+1} P_{J+1}(x)\,.
\end{align}

For the $+-$ helicity, the helicity amplitude with the total angular momentum $J\geq 1$ is
\begin{align}
    \mathcal{H}^{J}_{+-}
    &= \frac{c_R g^2}{4} \int_{0}^{\pi} d\theta\, \sin\theta\, d^J_{11}(\theta)
    \left(\DE + \DM\right) \left(1 + \cos\theta\right)\,, \notag\\
    &= \frac{c_R g^2}{2} \left[D_J^{\mathrm{E}} + D_J^{\mathrm{M}} + \frac{J+1}{2J+1}\left(D_{J-1}^{\mathrm{E}} + D_{J-1}^{\mathrm{M}}\right)
    +\frac{J}{2J+1}\left(D_{J+1}^{\mathrm{E}} + D_{J+1}^{\mathrm{M}}\right)\right]\,,
    \label{eq:HJ+-}
\end{align}
where we used the relation
\begin{align}
    (1+\cos\theta) d^J_{11}(\theta)
    &= \frac{J + 1}{2J + 1} P_{J-1}(\cos\theta) + P_J(\cos\theta)
    + \frac{J}{2J+1} P_{J+1}(\cos\theta)\,.
\end{align}
We note that $d^J_{11}(\cos\theta)$ vanishes for $J=0$.

\section{Decomposition of the helicity amplitude}
\label{sec:decompose}

As mentioned in Introduction, the Lorentz invariance is violated by the existence of the medium in QCD at finite $\mu$, so we only have to consider a frame in which the spin configuration of two incoming quarks is fixed.
In this setup, there is no mixing between the total spin $S$ and the orbital angular momentum $L$ as there is no $LS$-dependent term in the gluon interaction at leading order.
The orbital angular momentum purely arises from the relative motion of quarks, which is parameterized by the scattering angle $\theta$ between incoming and outgoing quarks;
through Eq.~\eqref{eq:DL}, this $\theta$ dependence is reflected exclusively in the partial wave expansion of the gluon propagator.
Therefore, the subscript of the gluon propagator becomes equal to the orbital angular momentum.

\begin{table}[t]
    \begin{tabular}{cccc|cc}
        \hline
        $C^{J0}_{J000}$ & $C^{J0}_{(J+1)010}$ & $C^{J0}_{J010}$ & $C^{J0}_{(J-1)010}$ & $C^{00}_{\frac12 \frac12 \frac12 (-\frac12)}$ & $C^{10}_{\frac12 \frac12 \frac12 (-\frac12)}$ \\ \hline
        $1$ & $-\sqrt{\frac{J+1}{2J+3}}$ & $0$ & $\sqrt{\frac{J}{2J-1}}$ & $\frac1{\sqrt{2}}$ & $\frac1{\sqrt{2}}$ \\ \hline
    \end{tabular}
    \begin{tabular}{cccc|cc}
        \hline
        $C^{J1}_{J001}$ & $C^{J1}_{(J+1)011}$ & $C^{J1}_{J011}$ & $C^{J1}_{(J-1)011}$ & $C^{01}_{\frac12 \frac12 \frac12 \frac12}$ & $C^{11}_{\frac12 \frac12 \frac12 \frac12}$ \\ \hline
        $0$ & $\sqrt{\frac{J}{2(2J+3)}}$ & $- \frac1{\sqrt{2}}$ & $\sqrt{\frac{J+1}{2(2J-1)}}$ & $0$ & $1$ \\ \hline
    \end{tabular}
    \caption{Clebsch-Gordan coefficients. \textit{Top:} for the $++$ channel. \textit{Bottom:} for the $+-$ channel.}
    \label{tab:CG}
\end{table}

Now, we further decompose the helicity amplitude in a fixed $J$ channel into channels with fixed values of $S$ and $L$.
Since each quark has a spin $1/2$, the total spin $S$ is restricted to either $0$ (singlet) or $1$ (triplet).
As in the nonrelativistic case, we classify using the term symbol $^{2S+1}L_J$.
One can do so by relating the two-particle states in the helicity basis $|JM;\lambda_1\lambda_2\rangle$ to the canonical state $|JM;LS\rangle$ labeled by the spin and orbital angular momentum through the following equation~\cite{Jacob:1959at,Chung:1971ri}:
\begin{align}
    |JM;\lambda_1\lambda_2 \rangle = \sum_{LS} \sqrt{\frac{2L+1}{2J+1}} C_{L0S\lambda}^{J\lambda} C_{\frac12 \lambda_1 \frac12 -\lambda_2}^{S\lambda} |JM;LS \rangle\,,
\end{align}
where $C^{JM}_{j_1 m_1 j_2 m_2} \equiv \langle j_1 m_1 j_2 m_2 | JM \rangle$ is the Clebsch-Gordan coefficient;
the relevant values are listed in Table~\ref{tab:CG}, which are taken from Sec. 8.5 in Ref.~\cite{Varshalovich:1988ifq}.
Below, we omit the magnetic quantum number $M$ in the ket as this is degenerate in the following analysis.
The helicity state $|J;++\rangle$ can be related to the canonical states as
\begin{align}
    |J;++ \rangle &= \frac1{\sqrt2} |J;L=J, S=0\rangle \notag\\
    &\quad + \sqrt{\frac{J}{2(2J+1)}} |J;L=J-1, S=1\rangle \notag\\
    &\quad - \sqrt{\frac{J+1}{2(2J+1)}} |J;L=J+1, S=1\rangle \,.
\end{align}
Note that $L=J$ state only has the spin singlet $S=0$ component.
Similarly, one can expand the helicity state $|J;+-\rangle$ as
\begin{align}
    |J;+- \rangle &= -\frac1{\sqrt2} |J;L=J, S=1\rangle \notag\\
    &\quad + \sqrt{\frac{J+1}{2(2J+1)}} |J;L=J-1, S=1\rangle \notag\\
    &\quad - \sqrt{\frac{J}{2(2J+1)}} |J;L=J+1, S=1\rangle \,.
\end{align}
As this helicity state does not contain a spin-singlet state, the total angular momentum has to be $J\geq 1$.

From these relation, one can project each helicity amplitude for a fixed total angular momentum $J$ onto channels with fixed $(S,L)$:
\begin{align}
    \mathcal{H}^J_{++} &= \frac12 \mathcal{H}^{S=0,L=J}_{++} + \frac{J}{2(2J+1)} \mathcal{H}^{S=1,L=J-1}_{++} + \frac{J+1}{2(2J+1)} \mathcal{H}^{S=1,L=J+1}_{++} \,,\\
    \mathcal{H}^J_{+-} &= \frac12 \mathcal{H}^{S=1,L=J}_{+-} + \frac{J+1}{2(2J+1)} \mathcal{H}^{S=1,L=J-1}_{+-} + \frac{J}{2(2J+1)} \mathcal{H}^{S=1,L=J+1}_{+-} \,.
\end{align}
By comparing these expressions to Eqs.~(\ref{eq:HJ++}, \ref{eq:HJ+-}), one can read out each helicity amplitude.
This is possible because, as mentioned above, the spin configuration is frozen due to the lack of the boost invariance, and the partial wave of the gluon propagator solely reflects the orbital angular momentum.
For the helicity amplitude in the $++$ channel with the total angular momentum $J$, the helicity amplitudes in the canonical $LS$ basis are
\begin{align}
    \mathcal{H}_{++}^{S=0,L=J} &= c_R g^2 \left(D_L^{\mathrm{E}} + 3D_L^{\mathrm{M}}\right)\,,\label{eq:Hsing++}\\
    \mathcal{H}_{++}^{S=1,L=J\pm 1} &= c_R g^2 \left(D_{L}^{\mathrm{E}} - D_{L}^{\mathrm{M}}\right)\,.
    \label{eq:Htrip++}
\end{align}
The helicity amplitude in the $+-$ channel with the total angular momentum $J\ge 1$, one can decompose as
\begin{align}
    \mathcal{H}_{+-}^{S=1,L} &= c_R g^2 \left(D_{L}^{\mathrm{E}} + D_{L}^{\mathrm{M}}\right)\,.
    \label{eq:H+-}
\end{align}
The expression is the same for $L = J-1$, $J$, and $J+1$.

Let us now discuss the Fermi statistics of the total wave function.
The total wave function should be antisymmetric.
When $J$ is even (odd), the spin and angular momentum wave function of the $(S,L)=(0,J)$ and $(1,J\pm1)$ channel is antisymmetric (symmetric), and the $(S,L)=(1,J)$ channel is symmetric (antisymmetric).
When color and flavor channels are symmetric in total, the remaining part of the wave function must be antisymmetric.
Therefore, when $J$ is even (odd) and the color-flavor wave function is antisymmetric (symmetric), $\mathcal{H}^J_{++}$ must vanish.
Similarly, when $J$ is even (odd) and the color-flavor wave function is antisymmetric (symmetric), only $\mathcal{H}^{S=1, L=J}_{+-}$ remains nonzero out of $\mathcal{H}^J_{+-}$.

\section{Classification of the pairing channel}
\label{sec:classify}

From these decomposed amplitudes, one can find the most attractive pairing channel for a given color and flavor representation of quarks.
This is possible because the RG equations in the effective theory near the Fermi surface, from which one can find the onset of the BCS instability, decouple in each channel labeled by the following quantum numbers:
\begin{align}
    \text{(color, flavor, helicity, ${}^{2S+1}L_J$)}\,.
    \label{eq:label}
\end{align}
The RG equation is formulated for a coupling function $G^0(t)$ and $G^i(t)$ of a four-point interaction in the effective theory near the Fermi surface at the scale $t = -\ln(\Lambda / \mD)$, where $\Lambda$ quantifies the energy difference relative to the Fermi surface.
At the scale $t=0$, the coupling functions $G^0$ and $G^i$ are matched onto $\DE$ and $\DM$~\eqref{eq:gluonprop} in QCD, respectively.

The RG equation in a channel with the helicity $++$ and $(S,L,J) = (0,L,L)$, which corresponds to the helicity amplitude in Eq.~\eqref{eq:Hsing++}, is~\cite{Fujimoto:2025inprep2}
\begin{align}
    \frac{d \left(G^{0}_{L++}+ 3 G^{i}_{L++}\right) }{dt} = -\frac{N\left(G^{0}_{L++} + 3 G^{i}_{L++} \right)^2}{2\left(1 + \frac{g^2}{9\pi^2}t\right)} + 3 \beta^i_{\mathrm{tree}}\,,
    \label{eq:RG1}
\end{align}
where $N = \mu^2/(2\pi^2)$ is the density of states on the Fermi surface, and $\beta^i_{\mathrm{tree}} = c_R g^2 / (6\mu^2)$ is the beta function arising from the renormalization of the tree-level magnetic amplitude.
The RG equation in a channel with the helicity $++$ and $(S,L,J) = (1,L,L\pm1)$ is
\begin{align}
    \frac{d \left(G^{0}_{L++} - G^{i}_{L++}\right)}{dt} = -\frac{N\left(G^{0}_{L++} - G^{i}_{L++}\right)^2}{6\left(1 + \frac{g^2}{9\pi^2}t\right)} - \beta^i_{\mathrm{tree}}\,.
    \label{eq:RG2}
\end{align}
This equation corresponds to the helicity amplitude in Eq.~\eqref{eq:Htrip++}.
Finally, 
The RG equation in a channel with the helicity $+-$ and $(S,L) = (1,L)$ is
\begin{align}
    \frac{d \left(G^{0}_{L+-} + G^{i}_{L+-}\right)}{dt} = -\frac{N\left(G^{0}_{L+-} + G^{i}_{L+-}\right)^2}{3\left(1 + \frac{g^2}{9\pi^2}t\right)} + \beta^i_{\mathrm{tree}}\,.
    \label{eq:RG3}
\end{align}
This equation holds for $J=L-1$, $L$, and $L+1$ and corresponds to the helicity amplitude in Eq.~\eqref{eq:H+-}.
These RG equation is common both for the color $\boldsymbol{3}$ and $\boldsymbol{6}$ representations, and the same for any flavor representation as well.
As one can see, in these equations, there is no mixing of the coupling function between different channels.
This guarantees that the pairing problem decouples between the system with different labels~\eqref{eq:label}.
Because of this decoupling, even if only one component of the interaction is attractive and all other repulsive, the system still undergoes a pairing instability into a state with the label~\eqref{eq:label} for which the interaction is attractive.

Let us find the most attractive pairing channel for a given color and flavor representation of a diquark, by resorting to the relations:
\begin{align}
    D_L^{\mathrm{E/M}} > D_{L+1}^{\mathrm{E/M}} > 0\,,\qquad
    D_L^{\mathrm{M}} > D_{L}^{\mathrm{E}}\,.
\end{align}
The first equation follows from the Rodrigues formula for the Legendre polynomial.
The second equation follows from that the electric interaction is Debye screened, while the magnetic interaction remains unscreened statically and is only screened dynamically, so the magnetic interaction dominates at a small angle $\theta \ll 1$.
These relations indicate that the most attractive pairing channel must be found in a channel with the smallest $J$ and $L$ possible.
The electric and magnetic propagators in $L=0$ and $1$ channels to leading order in $g$ are
\begin{align}
    \label{eq:DE0}
    \DE_0 &= \frac{1}{4\mu^2} \ln \left[1 + \left(\frac{2\mu}{\mD}\right)^2\right]\,,&
    \DE_1 &\simeq 
    \DE_0 - \frac{1}{2\mu^2}\,,\\
    \DM_0 &= \frac{1}{6\mu^2} \ln \left[1 + \frac{4}{\pi} \left( \frac{2 \mu}{\mD} \right)^3\right]\,,&
    \DM_1 &\simeq 
    \DM_0 - \frac{1}{2\mu^2}\,.
    \label{eq:DM1}
\end{align}

\begin{table}[t]
    \centering
    \begin{tabular}{c|cc|cccc}
        \hline
        Case & Color & Flavor & Helicity & $^{2S+1}L_J$ & Amplitude & Gap \\ \hline
        1 & $\boldsymbol{\bar{3}}$ & Antisymmetric & $++$ & $^1S_0$ & \eqref{eq:H1S0} & \eqref{eq:gap1S0}\\
        2 & $\boldsymbol{\bar{3}}$ & Symmetric & $++$ & $^1P_1$ & \eqref{eq:H1P1} & \eqref{eq:gap1P1} \\
        3 & $\boldsymbol{6}$ & Symmetric & $++$ & $^3P_0$ & \eqref{eq:H3P0} & \eqref{eq:gap3P0} \\
        4 & $\boldsymbol{6}$ & Antisymmetric & $++$ & $^3S_1$ & \eqref{eq:H3S1} & \eqref{eq:gap3S1} \\
        \hline
    \end{tabular}
    \caption{The most attractive channel for a given color and flavor representation.}
    \label{tab:gap}
\end{table}
We will consider each case with different color and flavor structure, and determine the remaining labels in Eq.~\eqref{eq:label}, namely, the helicity and ${}^{2S+1}L_J$;
in Table~\ref{tab:gap}, we summarize the most attractive channel for each case with a different color and flavor.
We compare the values of the pairing gap as a channel with the largest gap is favored the most.
For the pairing gap, we use the one evaluated by solving the RG equation (see Ref.~\cite{Fujimoto:2025inprep2} for details).
In all the cases, we find that the pairing between quarks with the same helicities are favored over the pairing between the opposite helicities.

\begin{figure}
    \centering
    \includegraphics[width=.49\columnwidth]{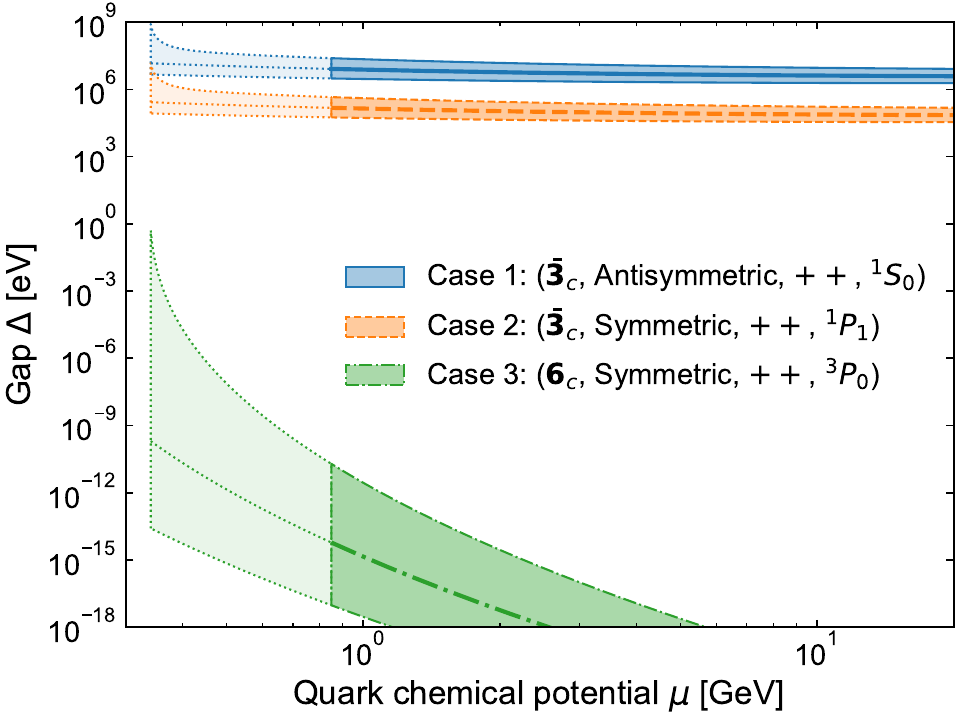}
    \includegraphics[width=.49\columnwidth]{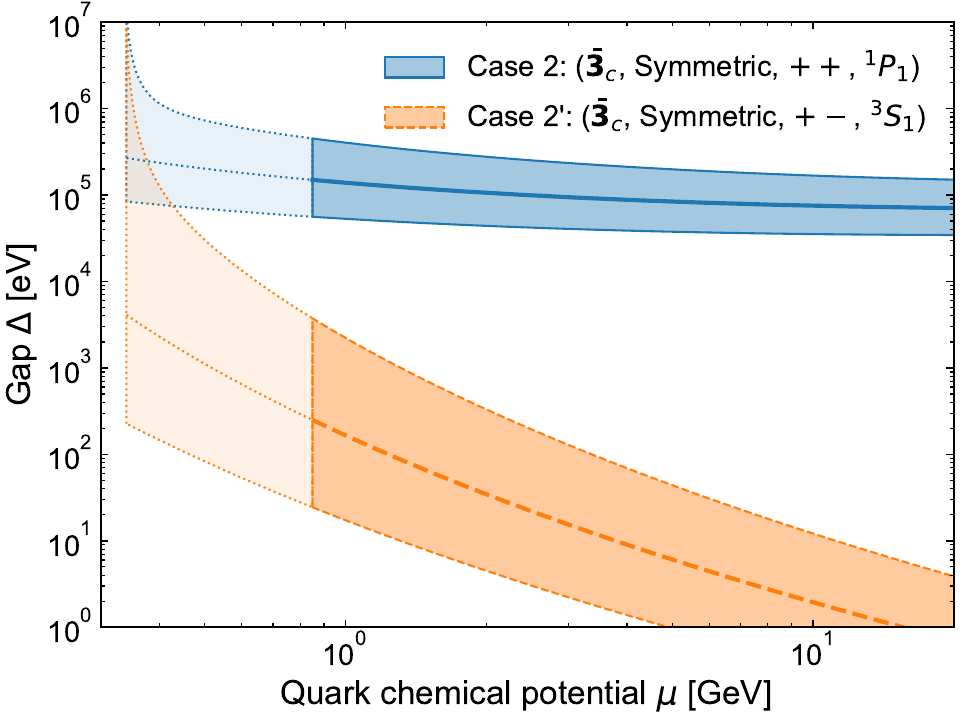}
    \caption{Pairing gap evaluated in the weak-coupling regime for $\Nf = 2$.
    The band corresponds to the renormalization scale variation.
    The dotted lines with lighter colors at small $\mu$ are the extrapolation to the region where perturbation theory may not be justified.
    \textit{Left:} For the different cases as classified in Table~\ref{tab:gap}.
    \textit{Right:} Comparison of the paring gaps~\eqref{eq:gap1P1} and \eqref{eq:gap3S1+-}.}
    \label{fig:gap}
\end{figure}
The magnitude of the gap as a function of quark chemical potential $\mu$ in $\Nf = 2$ is shown in the left panel of Figure~\ref{fig:gap}.
We use the running coupling constant from the two-loop QCD beta function evaluated in the $\overline{\mathrm{MS}}$ scheme with the scale $\Lambda_{\overline{\mathrm{MS}}} \simeq 340\,\text{MeV}$.
The band corresponds to the scale variation, for which we follow the convention by choosing the renormalization scale $\bar{\Lambda} = 2\mu$ and varying it by a factor two.

\subsection*{Case 1: Color $\boldsymbol{\bar{3}}$ and flavor antisymmetric}

We first consider the case that the color is in the antisymmetric $\boldsymbol{\bar{3}}$ representation and flavor is antisymmetric.
The remaining wave function has to be symmetric.
The most attractive channel is $J= L =0$, i.e., $^1 S_0$ channel in the same helicity scattering:
\begin{equation}
    \mathcal{H}_{++}^{{}^1S_0} = -\frac{2g^2}3 \left(D_0^{\mathrm{E}} + 3D_0^{\mathrm{M}}\right)\,.
    \label{eq:H1S0}
\end{equation}
The gap in this channel is
\begin{align}
    \Delta_{^1S_0} &\simeq e^{-\frac{\pi^2 + 4}{12}} \mD \left(1 + \frac{4\mu^2}{\mD^2}\right)^{\frac12}\left(1 + \frac{32\mu^3}{\pi\mD^3}\right) \exp\left(-\frac{\sqrt{3}\pi^2}{g}\right)\,.
    \label{eq:gap1S0}
\end{align}
As one can see in the left panel of Fig.~\ref{fig:gap} the gap in this channel stays almost constant with the value $\Delta_{^1S_0} \sim 5$--$50\,\text{MeV}$.

\subsection*{Case 2: Color $\boldsymbol{\bar{3}}$ and flavor symmetric}

Let us then consider the case that the color is in the $\boldsymbol{\bar{3}}$ representation and the flavor is in symmetric channel.
The remaining wave function has to be antisymmetric, and the smallest possible channels of $J$ and $L$ satisfying antisymmetry of the wave function are either in the $++$ helicity and $^1 P_1$ channel or in the $+-$ helicity and $^3 S_1$ channel.
The corresponding helicity amplitude for the former case is
\begin{equation}
    \mathcal{H}_{++}^{{}^1P_1} = - \frac{2g^2}3 \left(D_1^{\mathrm{E}} + 3D_1^{\mathrm{M}}\right)\,,
    \label{eq:H1P1}
\end{equation}
and for the latter case is
\begin{equation}
    \mathcal{H}_{+-}^{{}^3S_1} = - \frac{2g^2}3 \left(D_0^{\mathrm{E}} + D_0^{\mathrm{M}}\right)\,.
\end{equation}
From Eqs.~(\ref{eq:DE0}, \ref{eq:DM1}), it turns out that $\mathcal{H}_{++}^{{}^1P_1} < \mathcal{H}_{+-}^{{}^3S_1}$ when the coupling is $g \lesssim 0.92$, so the $^1 P_1$ channel in the same helicity scattering is the most attractive.
When the coupling constant is relatively large $g \gtrsim 0.92$, but still in the weak-coupling regime, the inequality becomes opposite: $\mathcal{H}_{++}^{{}^1P_1} > \mathcal{H}_{+-}^{{}^3S_1}$.
However, this does not directly infer that the ($+-$, $^3S_1$) channel is favored over the ($++$, $^1P_1$) channel when $g \gtrsim 0.92$;
the ($++$, $^1 P_1$) channel has stronger magnetic interaction compared to the ($+-$, $^3S_1$) channel as one can read out from the coefficient of the $\DM$ term.
The favored channel should be determined from the comparison of the gap.

One can then determine which is the most attractive channel by comparing the magnitude of the gap.
The pairing gap in these channels are
\begin{align}
    \Delta_{(++,{}^1P_1)} &\simeq e^{-4} e^{-\frac{\pi^2 + 4}{12}} \mD \left(1 + \frac{4\mu^2}{\mD^2}\right)^{\frac12}\left(1 + \frac{32\mu^3}{\pi\mD^3}\right) \exp\left(-\frac{\sqrt{3}\pi^2}{g}\right)\,,
    \label{eq:gap1P1} \\
    \Delta_{(+-,{}^3S_1)} &\simeq e^{-\frac{3(\pi^2 + 4)}{8}} \mD \left(1 + \frac{4\mu^2}{\mD^2}\right)^{\frac32}\left(1 + \frac{32\mu^3}{\pi\mD^3}\right) \exp\left(-\frac{3\sqrt{3}\pi^2}{\sqrt{2}g}\right)\,.
    \label{eq:gap3S1+-}
\end{align}
In the right panel of Fig.~\ref{fig:gap}, we plot the behavior of these gap as a function of $\mu$.
It turns out that the gap is always larger in the $^1 P_1$ channel compared to the $^3 S_1$ channel.

From Eqs.~(\ref{eq:gap1S0}, \ref{eq:gap1P1}), one can see that $\Delta_{^1S_0}$ and $\Delta_{^1P_1}$ has a fixed ratio:
\begin{align}
    \frac{\Delta_{^1P_1}}{\Delta_{^1S_0}} \simeq e^{-4} \simeq 0.018\,.
\end{align}
Therefore, the gap in the $^1P_1$ channel in the case 2 also stays constant with the value $\Delta_{^1P_1} \sim 0.1$--$1\,\text{MeV}$.

\subsection*{Case 3: Color $\boldsymbol{6}$ and flavor symmetric}

Next, we consider the case that the color is in the symmetric $\boldsymbol{6}$ representation and the flavor is in the symmetric channel.
This channel has been conventionally thought to be repulsive because $c_{\boldsymbol{6}} = 1/3 > 0$.
This is true in the vacuum, however, this is not the case in the medium.
The most attractive channel in the color $\boldsymbol{6}$ representation is $^3 P_0$ in the same helicity scattering:
\begin{equation}
    \mathcal{H}_{++}^{^3P_0} = - \frac{g^2}3 \left(D_{1}^{\mathrm{M}} - D_{1}^{\mathrm{E}}\right)\,.
    \label{eq:H3P0}
\end{equation}
We note that this contribution vanishes in the vacuum in which $\DE = \DM$.
They only survive when $\DE \neq \DM$, so this is purely the in-medium effect.
This is essentially why the same helicity pairing only accounted for the spin singlet pairing in the model employing the same interaction for the vacuum and in medium, see, e.g., Ref.~\cite{Alford:1997zt}.
We also note such an attraction is also present in QED, in which one replaces $c_{\boldsymbol{6}} = 1/3 \to 1$ and $g \to e$.
The corresponding gap is
\begin{align}
    \Delta_{{}^3P_0} \simeq e^{-\frac{3(\pi^2 + 4)}{2}} \mD \left(1 + \frac{4\mu^2}{\mD^2}\right)^{-\frac32}\left(1 + \frac{32\mu^3}{\pi\mD^3}\right) \exp\left(-\frac{3\sqrt{6}\pi^2}{g}\right)\,.
    \label{eq:gap3P0}
\end{align}
As one can see in Fig.~\ref{fig:gap}, this gap in the $^3P_0$ channel is less than $10^{-9}\,\text{eV}$ order in the weak coupling regime and decreases drastically as $\mu$ increases.
This pairing is stable against the splitting of the Fermi surface, so, there can in principle be superconductivity in this channel at absolute zero temperature.
In reality, the gap is too small and easily washed away by the thermal fluctuations, so this is negligible phenomenologically.
Even if one extrapolate this weak coupling results down to $\mu \sim 400\,\text{MeV}$, the gap can only reach up to the eV order.
Of course, this value should not be taken seriously at these low densities.
There may be an enhancement in the nonperturbative regime in this channel given the strong falling off behavior with increasing $\mu$, nevertheless, even with such enhancement, the gap is too small to be relevant.

\subsection*{Case 4: Color $\boldsymbol{6}$ and flavor antisymmetric}

Finally, we consider the color symmetric $\boldsymbol{6}$ and the flavor antisymmetric case.
Then, the most attractive channel is ${}^3S_1$ state in the same helicity scattering:
\begin{align}
    \mathcal{H}_{++}^{^3S_1} = - \frac{g^2}{3} \left(D_{0}^{\mathrm{M}} - D_{0}^{\mathrm{E}}\right) \,,
    \label{eq:H3S1}
\end{align}
which has the same magnitude as Eq.~\eqref{eq:H3P0} in the case 3.
\begin{align}
    \Delta_{{}^3S_1} \simeq e^{-\frac{3(\pi^2 + 4)}{2}} \mD \left(1 + \frac{4\mu^2}{\mD^2}\right)^{-\frac32}\left(1 + \frac{32\mu^3}{\pi\mD^3}\right) \exp\left(-\frac{3\sqrt{6}\pi^2}{g}\right)\,,
    \label{eq:gap3S1}
\end{align}
which is the same value as Eq.~\eqref{eq:gap3P0} in the case 3 up to the next-to-leading order in perturbation theory, so in Fig.~\ref{fig:gap}, we do not show this gap.

\section{Pairing patterns with the mismatched Fermi surface}
\label{sec:pattern}

\begin{figure}
    \centering
    \includegraphics[width=0.8\columnwidth]{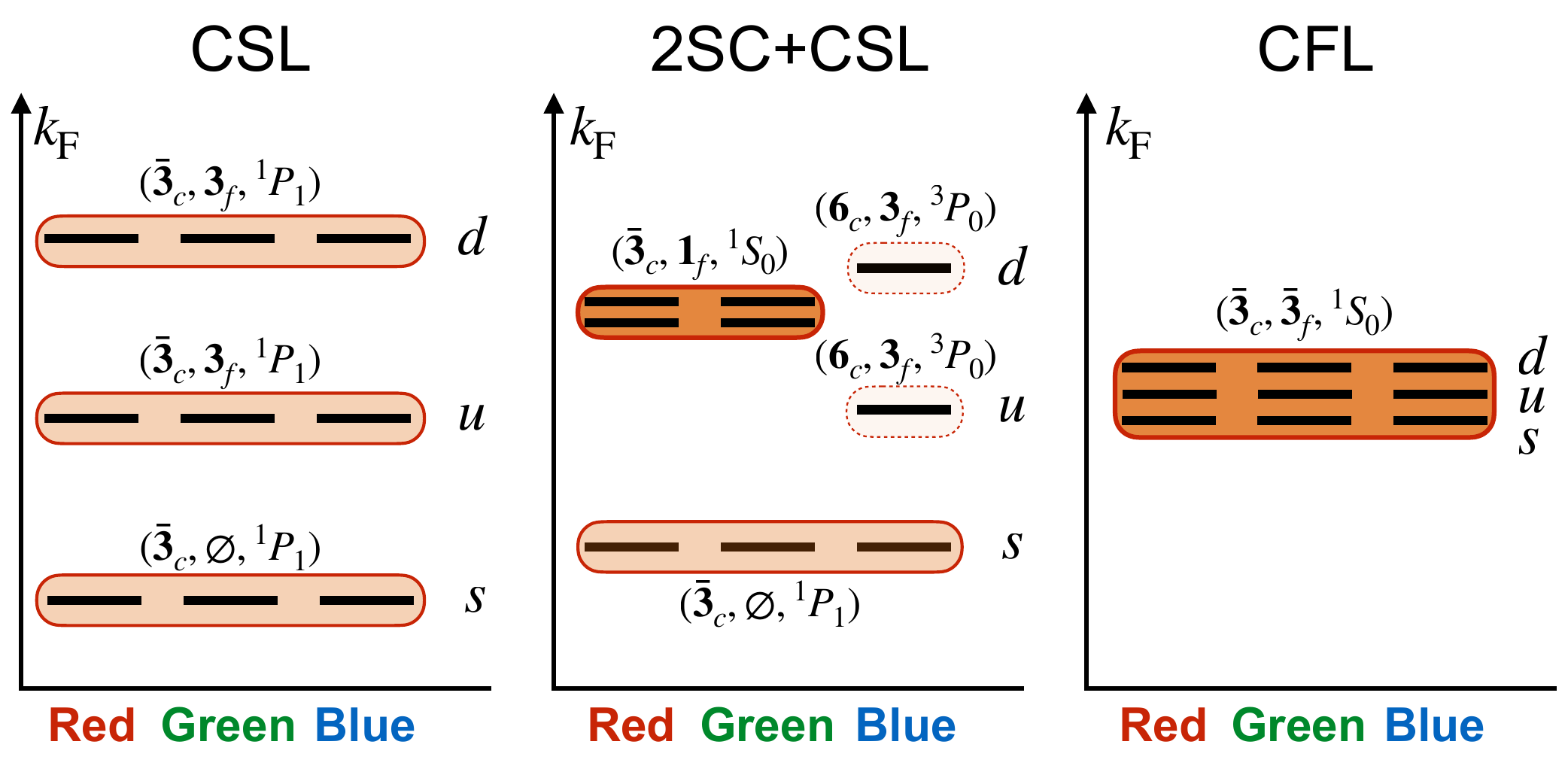}
    \caption{Illustration of the mismatched Fermi momenta of various colors and flavors of quarks with possible pairing patterns.
    The Fermi momentum splittings are not to the actual scale.
    Each pairing pattern are classified by (color, flavor, $^{2S+1}L_J$); the symbol $\varnothing$ in the strangeness sector refers to that the strange quarks are decoupled from the $SU(2)$ flavor symmetry, so the wave function is symmetric under the exchange of the flavor indices.
    The intensity of the color indicates the strength of the pairing.}
    \label{fig:pattern}
\end{figure}

In Fig.~\ref{fig:pattern}, we show the possible pairing patterns with mismatched Fermi momenta for $u$, $d$, and $s$ quarks in the weak-coupling regime.
The Fermi momenta for each quarks are not the same.
The separation is caused by the finite strange quark mass $m_s$, when simultaneously imposing the neutron star conditions, which are the beta equilibrium and electric charge neutrality condition.
The separation of the Fermi momenta is $\delta \kF \simeq m_s^2 / (4\mu)$.
As we lower the density, the strange quark mass is expected to become larger, so $\delta \kF$ is expected to be larger as well.
Figure~\ref{fig:pattern} is shown in the increasing density order from left to right.

\subsection{Three-flavor pairing}

Let us first turn to the rightmost panel in Fig.~\ref{fig:pattern}.
When $m_s$ is small, the flavor $SU(3)$ symmetry holds.
In terms of the flavor $SU(3)$, the representations of quarks decompose as $\boldsymbol{3} \otimes \boldsymbol{3} = \boldsymbol{\bar{3}} \oplus \boldsymbol{6}$.
In such a case, as classified in the previous section, the most attractive pairing occurs in the $(\boldsymbol{\bar{3}}_c, \boldsymbol{\bar{3}}_f, {}^1S_0)$ channel.
Because the color and flavor is in the same $\boldsymbol{\bar{3}}$ representations, all colors and flavors pair with each other, locking the color and flavor together (see Fig.~\ref{fig:pattern}, right panel)~\cite{Alford:1998mk}.

\begin{figure}
    \includegraphics[width=.49\columnwidth]{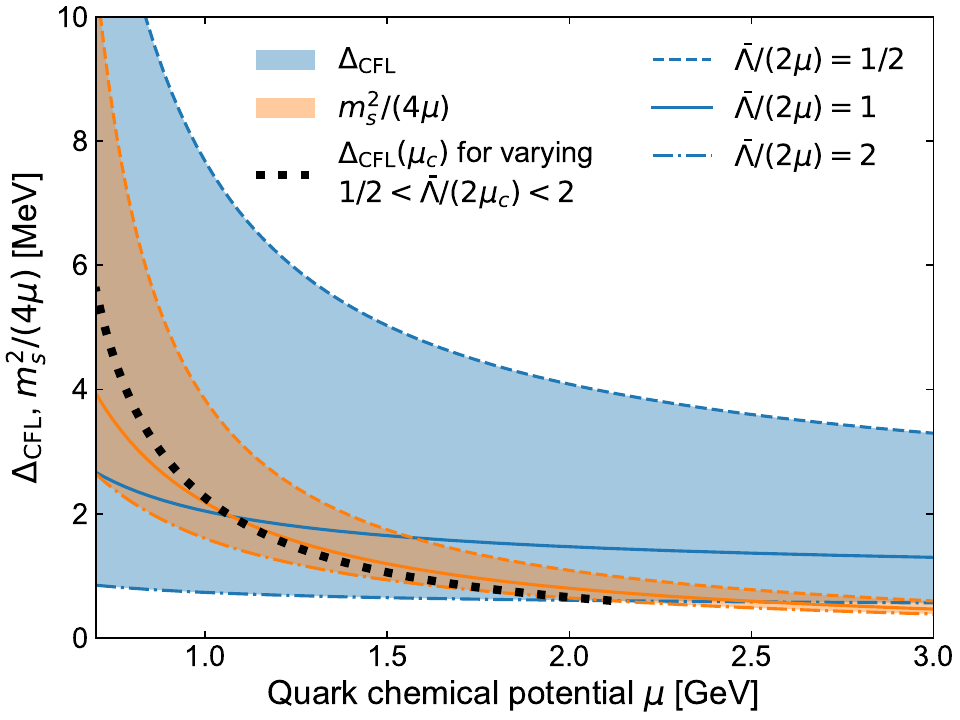}
    \includegraphics[width=.49\columnwidth]{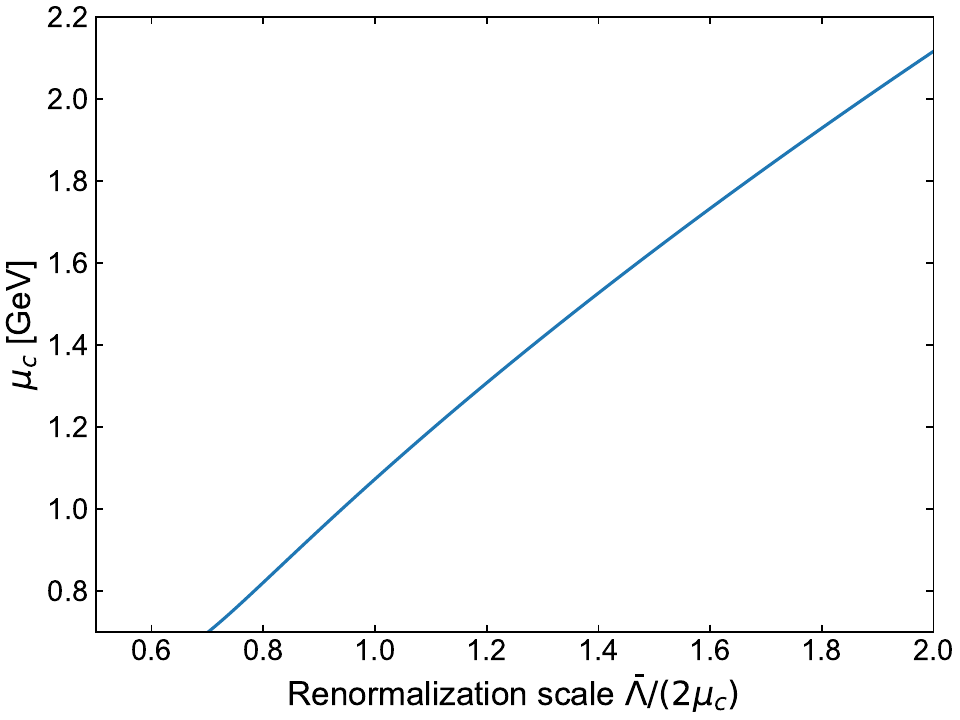}
    \caption{Comparison between the CFL gap $\Delta_{\rm CFL}$ and the stress on the Fermi energy induced by the strange quark mass $m_s^2 / (4\mu)$.
    \textit{Left:} Comparison between $\Delta_{\rm CFL}$ (shown by blue color) and $m_s^2 / (4\mu)$ (orange) with the scale variation uncertainty.
    Three representative values of the renormalization scale are chosen for either $\Delta_{\rm CFL}$ or $m_s^2/(4\mu)$, i.e., $\bar{\Lambda} / (2\mu) = 1/2$ (dashed line), $1$ (solid), $2$ (dash-dotted).
    Thick black dotted curve is the value of $\Delta_{\rm CFL}$ for a given $\mu_c$ with varying $\bar{\Lambda}$, which is shown on the right panel.
    \textit{Right:} Plot of $\mu_c$, at which $m_s^2(\bar{\Lambda})/(4\mu_c)$ becomes comparable to $\Delta_{\rm CFL} (\mu_c(\bar{\Lambda}), \bar{\Lambda})$, as a function of the renormalization scale $\bar{\Lambda} / (2\mu_c)$.}
    \label{fig:stress}
\end{figure}

All their Fermi momenta have a common value, and this is possible as long as the energy cost of forcing all species to have the same Fermi momentum is compensated by the Cooper pair condensation energy.
At the Chandrasekhar-Clogston point~\cite{1962ApPhL...1....7C,Clogston:1962zz}, at which $\delta \kF = \Delta / \sqrt{2}$, the pairing among the species with the Fermi surface disparity $\delta \kF$ undergoes a phase transition to the unpaired phase.
For the CFL case, this condition is $m_s^2 / (4\mu) \simeq \Delta_{\mathrm{CFL}}$~\cite{Rajagopal:2000ff,Alford:2001zr}.

In the left panel of Fig.~\ref{fig:stress}, we compare the CFL gap $\Delta_{\mathrm{CFL}}$ with the separation in the Fermi energy $m_s^2 / (4\mu)$.
For $m_s$, we use the two-loop running quark mass, following Ref.~\cite{Marquard:2015qpa}, with the value $m_s(\bar{\Lambda}=2\,\text{GeV})\simeq 93.5\,\text{MeV}$~\cite{ParticleDataGroup:2024cfk}.
We note that we do not include the factor $2^{-1/3}$ found in Ref.~\cite{Schafer:1999fe} in the CFL gap because the pairing channel does not include $\boldsymbol{6}$ channel, which seems to be the origin of this factor.
The band again corresponds to the renormalization scale variation, which is varied between $1/2 < \bar{\Lambda} / (2\mu) < 2$;
the upper bound corresponds to $\bar{\Lambda} / (2\mu) = 1/2$ and the lower bound to $\bar{\Lambda}/(2\mu) = 2$.
In the right panel of Fig.~\ref{fig:stress}, we plot the point $\mu_c(\bar{\Lambda})$ at which the CFL gap and the stress in the Fermi energy becomes comparable, i.e. $\Delta_{\mathrm{CFL}}(\mu_c(\bar{\Lambda}), \bar{\Lambda}) \simeq m_s^2 (\bar{\Lambda}) / (4\mu_c)$, for a given $\bar{\Lambda}$.
Physically, $\mu_c$ corresponds to the chemical potential at which the CFL phase becomes unstable against the unpaired phase.
In the left panel, we plot $\Delta_{\rm CFL}(\mu_c(\bar{\Lambda}), \bar{\Lambda})$ with a thick black dotted curve, which corresponds to the plot in the right panel.
What we can see from these figures is that the CFL phase can be unstable against the unpairing even at $\mu \gtrsim 1\,\text{GeV}$, depending on the value of $\bar{\Lambda}$.
So, even within the weak-coupling regime, transition from the pairing pattern shown in the rightmost panel in Fig.~\ref{fig:pattern} to the one shown in the middle or the leftmost panel can occur.
Nevertheless, there is a strong dependence on the renormalization scale, so the conclusive statement cannot be drawn at the moment.

\subsection{Two-flavor pairing}

Next we turn to the middle panel in Fig.~\ref{fig:pattern}.
When $\Nf = 2$, the representations of quarks in terms of the flavor $SU(2)$ decompose as $\boldsymbol{2} \otimes \boldsymbol{2} = \boldsymbol{1} \oplus \boldsymbol{3}$.
In this case, when the mismatch of the $u$- and $d$-quark Fermi momenta are moderately small, the most attractive pairing occurs in $(\boldsymbol{\bar{3}}_c, \boldsymbol{1}_f, {}^1S_0)$ channel according to the classification above [see $(\boldsymbol{\bar{3}}_c, \boldsymbol{1}_f, {}^1S_0)$ in Fig.~\ref{fig:pattern}, middle panel].
In this case, the two-flavor color superconductivity (2SC) only occurs among two out of three colors (e.g.\ red and green) for the flavor $\boldsymbol{1}$ channel in fixed-gauge description~\cite{Barrois:1979pv,Bailin:1983bm,Alford:1997zt,Rapp:1997zu}.
This is owing to the form of the condensate: 
\begin{align}
    \langle \psi^\top_\alpha C \gamma^5 \psi_\beta \rangle \propto \epsilon_{\alpha \beta 3} \Delta_{{}^1S_0}\,.    
\end{align}
The color indices $\alpha$ and $\beta$ in the condensate has to be antisymmetric, so the color direction of the gap matrix has to point to a specific direction (which is the third-direction in this case) in the fixed gauge description, and one can always gauge-rotate the gap orientation to a single direction in the color space.
In contrast, the flavor index is saturated as this is in the singlet channel, so the color index cannot be contracted with the flavor as in the CFL case.

Unpaired quarks in the remaining color (e.g.\ blue) in the symmetric $\boldsymbol{6}$ color representation can in principle undergo the pairing because there is a BCS instability in the $(\boldsymbol{6}_c, \boldsymbol{3}_f, {}^3P_0)$ channel, but the gap may be too small to be phenomenologically relevant.
The magnitude of this pairing gap may further be suppressed due to the Meissner effect from the 2SC pairing, as discussed in Ref.~\cite{Fujimoto:2025inprep2} (see also Ref.~\cite{Schafer:2006ue}).
Let us now discuss the form of the condensate for this pairing:
\begin{align}
    \langle \psi^\top_\alpha C \gamma^5 \gamma^i \nabla^j \psi_\beta \rangle \propto \delta_{\alpha 3} \delta_{\beta 3} \delta^{ij}  \Delta_{{}^3P_0}\,.
\end{align}
As this is in the spin-triplet and $p$-wave channel, this has spin dependence, which is characterized by $\gamma^i$ in the relativistic case, as well as dependence on orbital angular momentum, which is captured by the derivative $\nabla^i$.
The general index structure $\Upsilon^{ij} = \gamma^i \nabla^j$ can further be classified as follows~\cite{Fujimoto:2019sxg}:
\begin{align}
    ^3 P_0: \quad \Upsilon_{J=0} &= \Upsilon^{ii}\,,\\
    ^3 P_1: \quad \Upsilon^i_{J=1} &= \epsilon^{ijk} \Upsilon^{jk}\,,\\
    ^3 P_2: \quad \Upsilon^{ij}_{J=2} &= \Upsilon^{ij} - \frac13 \delta^{ij} \Upsilon_{J=0}\,.
\end{align}
As for the case of the $^3P_0$ channel, the operator is $\Upsilon^{ii}$, so the pairing is isotropic.

From symmetry perspectives, the pairing in this $(\boldsymbol{6}_c, \boldsymbol{3}_f, {}^3P_0)$ channel has an interesting consequence.
The symmetry breaking pattern becomes the same as in the nuclear phase~\cite{Fujimoto:2019sxg}.
Particularly, this phase with the 2SC condensate becomes superfluid owing to the breaking of the $U(1)_B$ symmetry, while the pure 2SC phase is not.
The quantum vortices associated with this superfluidity show peculiar properties~\cite{Fujimoto:2020dsa,Fujimoto:2021bes,Fujimoto:2021wsr};
they are similar to what is called the Alice strings~\cite{Schwarz:1982ec,Alford:1990mk}.

The remaining strange quark may undergo the single-flavor pairing, which we will turn in the next subsection.

\subsection{Single-flavor pairing}

Finally, we turn to the leftmost panel in Fig.~\ref{fig:pattern}.
When the separation of the $u$- and $d$-quark Fermi momenta are large, the 2SC pairing becomes unstable.
This is very likely in the current calculation because the 2SC and CFL pairing have the same magnitude of the gap.
There can still arise single-flavor color superconductivity (1SC), which has also been known as the spin-one color superconductor~\cite{Schafer:2000tw,Alford:2002rz,Buballa:2002wy,Schmitt:2002sc,Schmitt:2004et,Alford:2005yy}.
So, directly below the CFL density, the 1SC phase may be realized instead of the 2SC phase, i.e., the ground state changes directly from the rightmost panel in Fig.~\ref{fig:pattern} to the leftmost panel.

In the single-flavor pairing, the flavor wave function is always symmetric.
Among the flavor-symmetric ones, the $(\boldsymbol{\bar{3}}_c, {}^1P_1)$ channel is the most attractive, which was classified as the case 2 in the previous section.
Therefore, the 1SC pairing occurs in the ${}^1 P_1$ channel, which is a $p$-wave pairing rather than the triplet (spin-one) pairing.

Let us now discuss the form of the condensate.
Because this is the $p$-wave condensate, the gap can have the dependence on the relative motion of quarks $\nabla^i$:
\begin{align}
    \langle \psi^\top_\alpha C \gamma^5 \nabla^i \psi_\beta \rangle \propto \epsilon_{\alpha \beta \gamma} \phi^{\gamma i}\,,
\end{align}
and in this case, one can take $\phi^{\gamma i} = \delta^{\gamma i}\Delta_{{}^1P_1}$ because the color and real space both have three directions.
This is known as color-spin locking (CSL), although in this case the color is locked with orbital angular momentum to be precise.
In this case, the condensate is isotropic both in the color space and the real space.
The RG equation alone cannot determine the form of the condensate, but it can determine the most attractive interaction channel from the BCS instability located by increasing the scale parameter from $t=0$.
Whichever hits the first with the smallest $t$ is the most attractive channel.
In this case, for quarks with any colors, the BCS instability is hit first in the $(\boldsymbol{\bar{3}}_c, {}^1P_1)$ channel.
The CSL pairing pattern can allow quarks with every color to pair in this channel.
Otherwise, it has to point to a specific direction as in the 2SC pairing, only allowing quarks with two colors out of three pairing up.
This is in agreement with the preceding studies, in which authors find that the CSL pairing occurs in the ground state for the 1SC phase~\cite{Schafer:2000tw,Schmitt:2004et}.
However, we have found the CSL pairing occurs between quarks with the same helicity in contrast to the preceding literature, in which they found the pairing in the opposite helicity channel.

Although the gap is suppressed compared to the 2SC phase, this 1SC pairing remains stable against the separation of the Fermi surface.
The gap we found in the weak-coupling regime is significantly larger than what has been known.
This single-flavor pairing can be important phenomenologically, as was explored, e.g., in Refs.~\cite{Schmitt:2003xq,Wang:2009if,Wang:2010ydb}, because all the quarks are gapped.

\section{Conclusions}
\label{sec:conclusions}

We have calculated the one-gluon exchange (OGE) helicity amplitudes in fixed total angular momentum $J$ channels, and classified them in a nonrelativistic manner.
This classification is justified by the fact that, in a Lorentz-noninvariant medium, the OGE interaction can be decomposed into distinct angular momentum channels labeled by $(S,L)$, and that the corresponding renormalization group (RG) equations decouple at leading order.
While angular-momentum structures have been considered in previous RG analyses (e.g., Ref.~\cite{Hsu:1999mp}), certain aspects of their interpretation remained unclear.

From the helicity amplitude analysis, we found that, in addition to the vacuum-attractive color $\boldsymbol{\bar{3}}$ channel, the vacuum-repulsive color $\boldsymbol{6}$ channel can also become attractive due to medium effects.
This feature, described in Eq.~\eqref{eq:Htrip++}, occurs exclusively in the spin-triplet sector.

Based on the OGE interaction, we then examined, for a given color-flavor representation, which (helicity, $^{2S+1} L_J$) combination is most attractive for Cooper pairing.
For the antisymmetric flavor channel with color $\boldsymbol{\bar{3}}$, corresponding to the standard CFL and 2SC phases, pairing occurs in the ${}^1S_0$ channel, in agreement with conventional understanding.
For the symmetric flavor channel, relevant to single-flavor pairing, the same-helicity ${}^1P_1$ and the opposite-helicity ${}^3S_1$ channels can in principle compete;
however, our results indicate that the former is significantly more attractive.
This conclusion deviates from the conventional picture of single-flavor pairing.
Our finding here may also have a consequence for the topological properties of the single-flavor pairing~\cite{Sogabe:2024yfl}.

We also investigated the pairing patterns in the presence of Fermi-surface mismatches.
The ${}^1S_0$ channel remains the most stable, favoring the CFL phase for three flavors.
For two flavors, the system realizes the 2SC phase, leaving blue quarks unpaired.
Although the BCS instability persists in this channel, the resulting gap is too small.
Our analysis was restricted to the massless case; in the massive, nonrelativistic regime, the $LS$ interaction is known---via the Fermi-Breit expansion---to render this channel attractive, suggesting that if the 2SC phase persists into the nonrelativistic, nonperturbative region, the blue quarks might also form pairs~\cite{Fujimoto:2019sxg};
The pairing mechanism is similar to the ${}^3P_2$ superfluidity in neutron star matter~\cite{Hoffberg:1970vqj,Tamagaki:1970ptp}.
For the single-flavor case, our results are consistent with the color-spin locked (CSL) phase.
However, we did not resolve the degeneracy with respect to the magnetic quantum number $M$ associated with total $J$.
The tree-level interaction is insensitive to $M$, but other physical mechanisms could lift this degeneracy.

From the viewpoint of quark-hadron continuity~\cite{Schafer:1998ef}, both the CSL and the 2SC phase supplemented by blue quark pairing break the same $U(1)_B$ symmetry as in nuclear matter.
With respect to chiral symmetry, while we did not analyze it in detail here, the CSL phase may break chiral symmetry~\cite{Schafer:2000tw}, thereby enabling quark-hadron continuity.
In the 2SC + blue quark condensate phase, chiral symmetry breaking is not guaranteed but could occur under certain conditions~\cite{Fujimoto:2019sxg}.

\begin{acknowledgments}
I would like to thank Sanjay~Reddy for a suggestion on the manuscript.
This research was supported in part by the Japan Science and Technology Agency (JST) as part of Adopting Sustainable Partnerships for Innovative Research Ecosystem (ASPIRE) Grant No.\ JPMJAP2318, the National Science Foundation Grant No.\ PHY-2020275, and the Heising-Simons Foundation Grant 2017-228.
\end{acknowledgments}

\bibliography{pairing}

\end{document}